# A wavelength-tunable mode-locked Yb-doped fiber laser based on nonlinear Kerr beam clean-up effect

**Shanchao Ma (马善超)[1], Baofu Zhang (张宝夫)[2]*, Qiurun He (何秋润)[1], Jing Guo(郭靖)[3], and Zhongxing Jiao (焦中兴)[1]**

[1] School of Physics, Sun Yat-sen University, Guangzhou 510275, China

[2] School of Materials Science and Engineering, Dongguan University of Technology, Dongguan 523808, China

[3] School of Opto-Electronics, Beijing Institute of Technology, Beijing 100081, China

*Corresponding author: zhangbf5@mail.sysu.edu.cn; ** corresponding author: jiaozhx@mail.sysu.edu.cn

We demonstrate a novel approach to achieve wavelength-tunable ultrashort pulses from an all-fiber mode-locked laser with a saturable absorber based on nonlinear Kerr beam clean-up effect. This saturable absorber was formed by a single-mode fiber spliced to a graded-index multimode fiber, and its tunable band-pass filter effect is studied by a numerical model. By adjusting the bending condition of graded-index multimode fiber, the laser could produce dissipative soliton pulses with their central wavelength tunable from 1040 nm to 1063 nm. The pulse width of output laser could be compressed externally to 791 fs, and the signal to noise ratio of its radio frequency spectrum was measured to be 75.5 dB. This work provides a new idea for the design of wavelength-tunable mode-locked laser which can be used as an attractive light source in communication system.

## 1. Introduction

Wavelength tunable lasers are widely used in optical communication, detection, and remote sensing. Compared to tunable continuous-wave lasers, tunable mode-locked lasers can generate ultra-short pulses with high peak power and wide bandwidth that enable many important advances, such as high-sensitivity optical absorption measurement, multi-photon microscopy, and super-continuum light source used in dense wavelength division multiplexing. Moreover, due to the high conversion efficiency, low temperature sensitivity and compact structure of fiber lasers, tunable mode-locked fiber lasers [1-4] have attracted extensive research interest.

Up to date, there are several typical approaches to achieve the tunability of mode-locked fiber lasers. The most common method is to add a tunable spectral filter into the laser cavity [1-4]. Among them, although chirped fiber Bragg gratings have been used to achieve wavelength tuning [1], the tuning range is limited to a few nanometers. Fourier domain programmable optical processor can also be used as spectral filter [2], but it was based on a two-dimensional liquid-crystal-on silicon array which cannot be easy to construct. Multimode interference (MMI) based band-pass filters can also be used to achieve tunable output [3,4], but the mode-locking operation based on nonlinear MMI was very sensitive to the length of graded-index multimode fiber (GIMF). Therefore, in order to meet the requirements of emerging applications, more attention should be paid to find out a simple and reliable solution for wavelength-tunable mode-locked fiber lasers.

Nonlinear Kerr beam cleanup (NL-KBC) effect occurs when high-power laser propagates in the GIMF. With the increasing laser power or longer fiber length, energy of high-order modes (HOMs) can be irreversibly coupled into the fundamental mode, where nonlinear Kerr effect is the driving mechanism [5, 6]. Based on this NL-KBC effect, we have constructed all-fiber saturable absorbers (SAs) with the structure of a long GIMF spliced to a short single-mode fiber (SMF) [7, 8]. We experimentally investigated characteristics of these SAs [7], and then demonstrated a mode–locked ytterbium-doped fiber laser using one of them [8]. We found that the bending of GIMF in this laser might lead to different transverse mode distribution of the beam with different wavelengths, resulting in a band-pass filter characteristic along with the NL-KBC effect. This may be a new approach to achieve wavelength-tunable output in the mode-locked fiber lasers.

In this paper, an all-fiber wavelength-tunable mode-locked laser based on NL-KBC effect is demonstrated, and its tuning mechanism is theoretically studied by a numerical model. For this laser, mode-locking operation with a wide tunable range from 1040 nm to 1063 nm can be obtained; the output pulse duration can be compressed externally to 791 fs.

## 2. Theoretical model and numerical simulation

In this section, we will investigate the band-pass filter effect and saturable absorption property of an SA based on NL-KBC by a numerical model. In this model, the SA can be divided into four parts which are centrally aligned (shown in Fig. 1): an input SMF, a short GIMF segment with a small radius of curvature, a long and straight GIMF segment, and an output SMF. A commercial GIMF (Nufern, GR-50/125-

23HTA) is used here; both input and output SMFs have the same type (Nufern, Hi1060-XP). It is assumed that only the multimode-coupling effect introduced by the bending condition is considered in the short GIMF segment, while dispersion and nonlinear effects are considered in the long GIMF segment. The bending condition here refers to the curved length and curvature radius. Since the mode field diameter of this SMF is close to the fundamental-mode (FM) diameter of GIMF, we can obtain approximate solutions by assuming that all laser energy from the input SMF will be transferred to the excited FM energy in the GIMF, while only the FM energy at the end of GIMF can be coupled into the output SMF. With this approximation, the transmittance of this SA can be defined as the ratio of FM energy at the end of GIMF to the launched laser energy from the input SMF. The main theoretical method and some important assumptions are listed at the flowchart in Fig. 1(b).

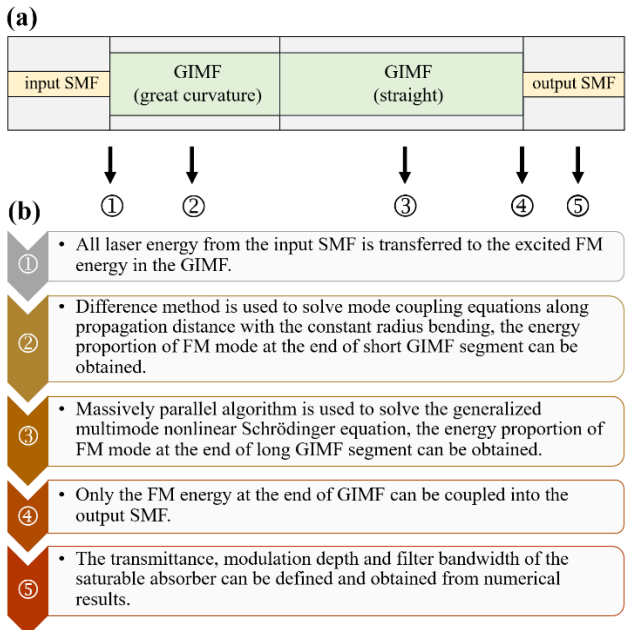

**Fig. 1.** The structure of a saturable absorber based on NL-KBC.

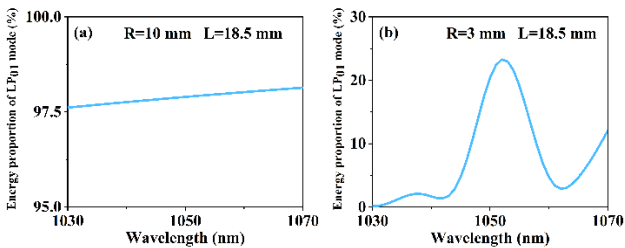

**Fig. 2.** The simulated proportion of LP01 mode at the end of the short graded-index multimode fiber segment with different bending conditions. R: radius of the curvature; L: the curved length.

In the short GIMF segment, the influences of bending condition on the transverse mode distribution of laser beam with different wavelengths are studied. For qualitative analysis, the bending of GIMF can greatly affect mode propagation constants and hence the coupling coefficients between laser transverse modes [9]. Besides, mode propagation constants also depend on the laser wavelength. Therefore, when a FM laser beam propagates through a curved GIMF, HOMs can be excited; the bending condition and laser wavelength will have great influences on the output transverse mode distribution. In order to figure out these influences, numerical simulations are performed by solving the wave coupling equation through a curved GIMF; this numerical method is similar to Sohn's work [10]. First, the transverse eigenmodes and their propagation constants of GIMF can be obtained by analytically solving the waveguide eigen equation. Then, local transverse mode fields are matched at a succession of infinitesimal comer bends to calculate mode coupling induced by the constant radius bending. Finally, the energy proportion of different modes versus laser wavelength, propagation distance or curvature radius, can be obtained at the end of GIMF. The detail of three steps above can be found in Page 10-11 and 35-46 of Ref. [10]. In this simulation, the length of short GIMF segment is set to be 40 mm with its curved length of 10-40 mm and a curvature radius of 2-10 mm. Moreover, it is assumed that the propagating laser beam is monochromatic, and only linear-polarized (LP) modes are considered. Since the energy proportion of LP01 mode at the end of short GIMF segment has great influences on the transmittance of our SA (detail in the following paragraph), this energy proportion with different bending conditions is shown in Fig. 2. For specific condition, the energy proportion of LP01 mode varies with the laser wavelength. These numerical results are in good agreement with the qualitative analysis mentioned above. As a result, the transverse mode distribution versus wavelength at the end of short GIMF segment will be served as the input condition for numerical simulations in the long GIMF segment.

In the long GIMF segment, the influences of input condition on the output laser characteristics are studied. Since the main issue here is multimode nonlinear optical pulse propagation, numerical simulations are performed by solving a generalized multimode nonlinear Schrödinger equation using massively parallel algorithm (MPA). The MPA numerical solver we used is the same as that reported by L. G. Wright et al. [11], and this solver edited in MATLAB is available online [12]. In our numerical model, only the first ten LP modes are considered. Moreover, the effects of self-phase modulation, stimulated Raman scattering, and the first five order dispersion are included in the simulations, but the random linear distortions such as random radius perturbation and random index perturbation introduced by the imperfections or bending of GIMF are not included. The length of long GIMF segment is set to be 2 m, and the input laser pulse has a duration of 50 fs with different pulse energy and central wavelength. For the situation of short GIMF segment with 18.5-mm curved length and 3-mm curvature radius, as shown in Fig. 2(b), the energy proportion of LP01 mode versus wavelength is simple and clear with a single peak at 1052 nm. Therefore, in the following simulations, this situation is chosen as the input condition of long GIMF segment. First, high-energy multimode laser pulse propagation through the long GIMF segment is studied. When the laser pulse energy is set to be 50 nJ and its wavelength is 1052 nm, the evolution of laser transverse mode distribution along the GIMF is shown in Fig. 3(a). During the propagation, the proportion of LP01 mode first increases from 23% to 52%, and then gradually keeps stable at 40%; most of HOMs have a lower proportion at the end of GIMF. Therefore, part of HOM energy is irreversibly coupled into the fundamental mode, which indicates the NL-KBC effect occurs in this case. Then, we focus on the optical properties of this SA which depend on output laser characteristics after it propagates through this long GIMF segment. The relationship between transmittance of the SA and laser wavelength is presented in Fig. 3(b); the cases of input laser beam with high pulse energy (50 nJ) and low pulse energy (0.001 nJ) are both considered. As shown in Fig. 3(b), the transmittance of SA versus wavelength have a similar shape with the energy proportion of LP01 mode as shown in Fig. 2(b), which results in a band-pass filtering effect. This band-pass filter can be tunable since its central wavelength and bandwidth can be adjusted by changing the bending condition of short GIMF segment and hence the input transverse mode distribution of long GIMF segment. In addition, the transmittance with higher input pulse energy (red line) is higher than that with lower energy (blue line) at each wavelength, which confirms the saturable absorption effect of this SA.

Finally, in order to further evaluate optical properties of the SA, its transmittance under different input pulse intensity is studied. The numerical results are presented in Fig. 3(c); the laser wavelength is chosen to be 1052 nm which is the central wavelength of band-pass filter effect. The transmittance first grows monotonically as the input pulse energy increased and then decrease when the pulse intensity is high The decrease of transmittance may be caused by stimulated Raman scattering (SRS) which can lead to the serious dissipation of propagating laser power [6]. This assumption is also supported by the monotonically increasing transmittance when SRS is not included in the simulation. In order to evaluate the modulation depth and saturation intensity of this

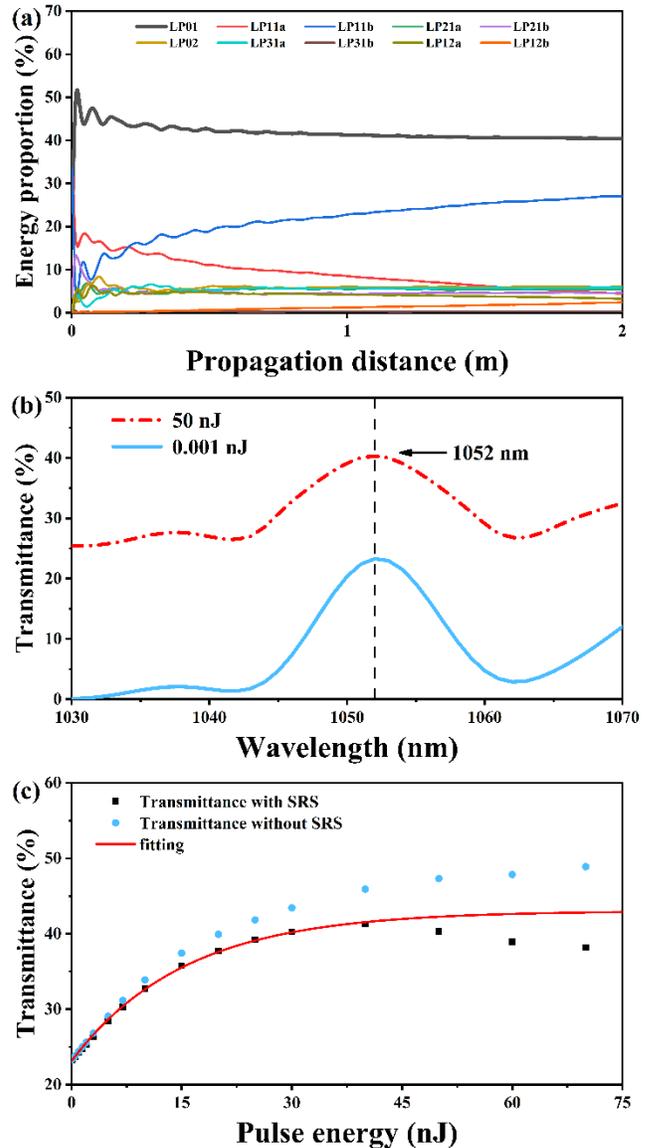

Fig. 3. Numerical results for the saturable absorber: (a) the evolution of modal energy proportion along the long GIMF segment when the laser wavelength is 1052 nm; (b) the transmittance of saturable absorber versus input laser wavelength when the pulse energy is 50 nJ (red line) and 0.001nJ (blue line); (c) the transmittance of saturable absorber versus input laser intensity when the laser wavelength is 1052 nm

SA, the transmittance curve before the decrease can be fitted by

$$T = T_0 + q \cdot \left[1 - \exp\left(-\frac{I}{I_{SAT}}\right)\right] \qquad (1)$$

where $T$ is the transmittance, $T_0$ is the initial transmittance, $q$ can be defined as the modulation depth, $I$ and $I_{SAT}$ are the input and saturation intensity, respectively. As shown in Fig. 3(c), the curve fits well with the numerical results; the initial transmittance, modulation depth and saturation intensity are fitted to be 23.12 %, 19.91 % and 15.41nJ, respectively. These results further ensure the saturable absorption effect of the SA.

In conclusion of this section, a numerical model is used to study the optical properties of the SA based on NL-KBC effect. The simulation results indicate that the bending condition plays an important role in the transverse mode distribution of GIMF and hence the transmittance of SA. Saturable absorption and band-pass filter effect with tunable central wavelength of the SA can be achieved with specific bending condition of GIMF when the NL-KBC occurs. Therefore, this SA may be promising for constructing tunable mode-locked lasers.

### 3. Experiments and results

In order to verify the tunable band-pass effect of the SA based on NL-KBC, we built up a mode-locked ytterbium-doped fiber laser with this SA. As shown in Fig. 4, the configuration of the laser is similar to that demonstrated in our previous work [8] except the polarization independent isolator (PI-ISO) used in the cavity had a wide bandwidth (>50 nm). In this laser, a 976-nm laser diode was used as the pump source, and its pump power was coupled into an ytterbium-doped fiber (Yb1200-4/125, LIEKKI) through a commercial wavelength division multiplexer. The length of this gain fiber was chosen to be 0.3 m in order to reduce the mode-locking threshold and increase the output power. 10 % of the mode-locked laser was output from a coupler after the ytterbium-doped fiber. The output port of coupler was connected with an isolator to prevent the influence of reflected light. The function of PI-ISO was to ensure the unidirectional operation of laser and eliminate the possibility of nonlinear–polarization-evolution mode locking. Two polarization controllers (PCs) were used to optimize intracavity birefringence in order to achieve stable mode-locking state. The SA was the same as what we mentioned in the simulation section, and it was placed between the two PCs. The bending condition of short GIMF segment can be adjusted by two XY-axis translation stages with resolution of 10 µm. All single-mode passive fibers used in this laser were the same type (HI-1060, Corning); the total cavity length was measured to be about 7.4 m.

When the short GIMF segment was adjusted to a specific bending condition, the self-starting mode-locking operation could be achieved with only increasing the pump power. When the operating central wavelength was tuned to 1040 nm, the single-pulse mode-locking state was obtained at the threshold of 60 mW. With the pump power increasing to 160 mW, mode-locking state could be maintained and its output pulse characteristics was measured.

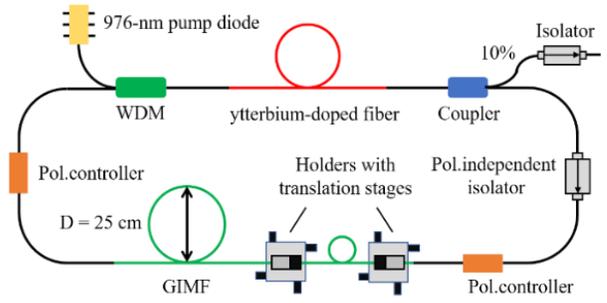

**Fig. 4**. Experimental setup of the tunable mode-locked ytterbium-doped fiber laser based on NL-KBC. WDM: wavelength division multiplexer; GIMF: graded-index multimode fiber; Pol.: polarization; D: diameter.

The output pulse trace and radio-frequency (RF) spectrum are presented in Fig. 5. The period of laser pulses was about 36 ns, and the first peak of RF spectrum was centered at 27.566 MHz, corresponding to the cavity length of 7.4 m. The signal-to-noise ratio of this RF spectrum was measured to be 75.5 dB; no multi-pulse or harmonic mode-locked signals were found in RF spectrum. These results further confirmed that the laser was operating at a single-pulse mode-locking state. Fig. 5(c) shows the laser output spectrum at the same pump power of 160 mW; its central wavelength was 1040 nm with a 20-dB bandwidth of 6.2 nm. The spectrum presented a cat-ear-like trace, which is a typical characteristic of dissipative soliton from all-normal-dispersion fiber lasers. The autocorrelation trace of output pulse was also measured (see Fig. 5(d)); assuming a Gaussian pulse shape, the pulse duration was calculated to 25.7 ps. It could be compressed externally to 791 fs using a single 1200-line/mm grating.

In addition, wavelength-tunable mode-locking operation was studied in this laser. By adjusting the translation stages, different bending condition of GIMF can be achieved. As a result, the central wavelength of output laser could be tuned from 1040 nm to 1063 nm while the mode-locking state could be maintained. The output spectra with different central wavelength are presented in Fig. 6. They all showed a cat-ear-like trace, but their mode-locking thresholds are different ranged from 60 mW to 330 mW. The spectral tuning range of the mode-locking operation may be limited by the gain bandwidth of the ytterbium-doped fiber or band-pass filter effect induced by other fiber devices. The phenomena of tunable laser output can be inferred by the simulation results, which verified the validity of our numerical model in Section 2.

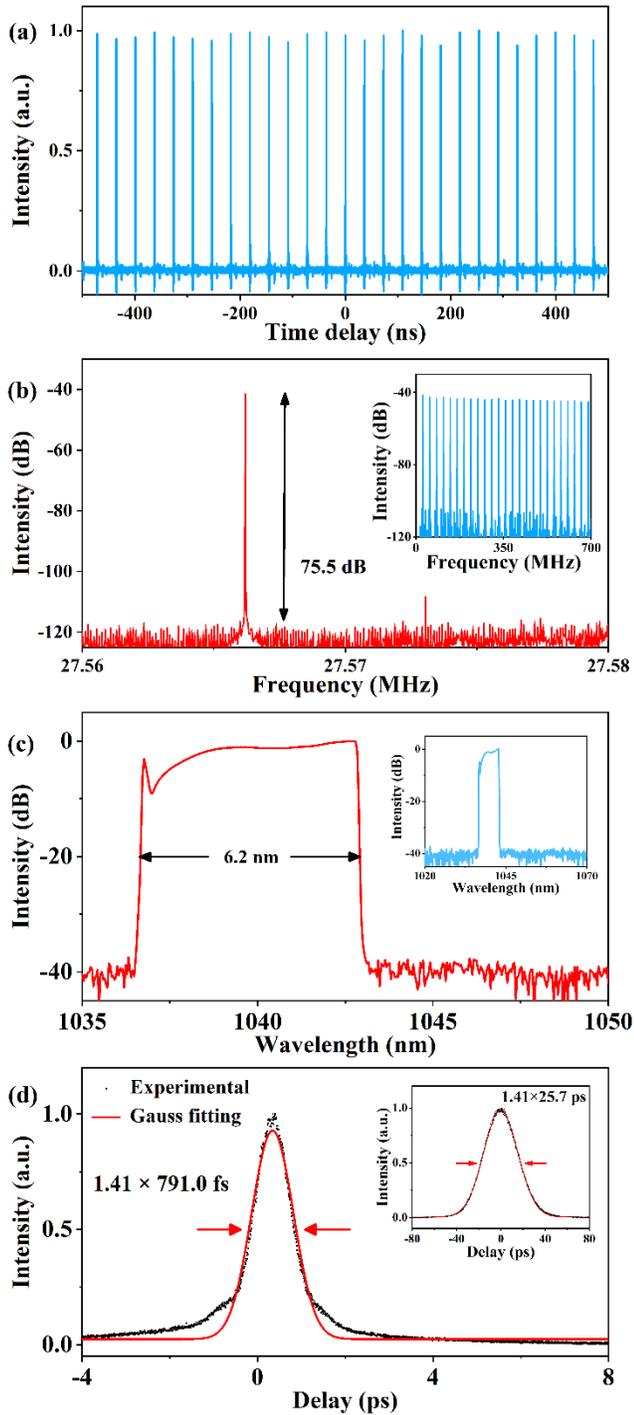

**Fig. 5.** (a) Single-pulse mode-locked pulse train when the pump power was 160 mW. (b) The output radio frequency spectrum with 1 Hz resolution bandwidth. Inset: The output radio frequency spectrum with 100 Hz resolution bandwidth. (c) The output optical spectrum with a spectral resolution of 0.04 nm; inset: the output optical spectrum with a spectral resolution of 0.1 nm. (d) The autocorrelation trace of dechirped output pulses with Gaussian fitting; inset: the autocorrelation trace of output pulses with Gaussian fitting.

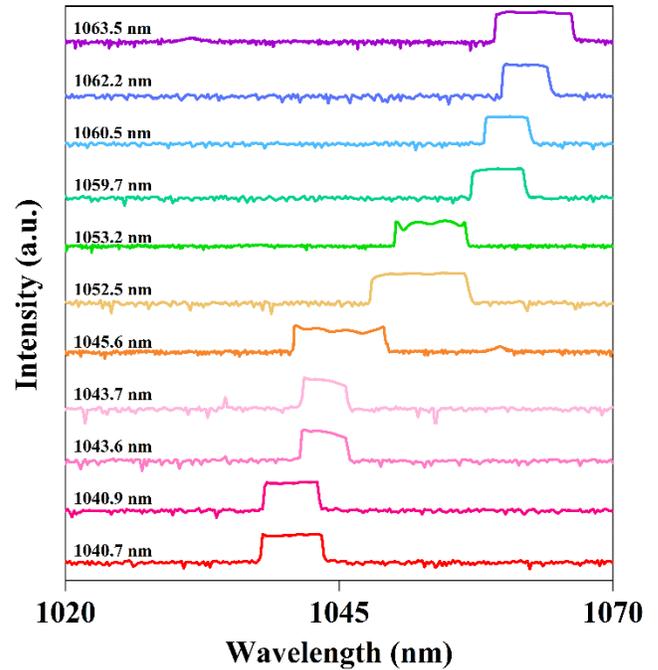

**Fig. 6.** Output spectra of the tunable mode-locked fiber laser with central wavelength ranged from 1040.7 nm to 1063.5 nm.

## 4. Conclusion

We have demonstrated a novel all-fiber tunable mode-locked laser with a SA based on NL-KBC effect. The mechanisms of tunable band-pass filter and saturable absorption effects in the SA have been studied by an original numerical model. Guided by the simulation results with adjusting the bending condition of GIMF, the laser could produce stable dissipative soliton pulses with tunable central wavelength ranged from 1033 nm to 1063 nm. This work provides a novel approach for constructing wavelength tunable mode-locked fiber lasers which can be used in optical communication, optical sensing and other important fields. Future works will focus on the all-polarization-maintaing-fiber configuration and precise tuning of the mode-locked wavelength in these lasers.

**Acknowledgment**

The authors would like to acknowledge Zhanwei Liu for the helpful discussion on the numerical model and the MPA program. We acknowledge Ugur Tegin for helpful discussion on the laser design, and Yu Duan of Emgo-tech Ltd (Zhuhai), Fujuan Wang, Jiaoyang Li for the technical support in this work.